%
\magnification\magstep1
\def\np{\hfill\vfill\eject\noindent}    
\def\smallskip{\vskip 3pt}
\def\medskip{\vskip 10pt}
\def\bigskip{\vskip 24pt}
\def\sol{$_\odot$ }                     
%
%
\def\ltsimeq{\,\raise 0.3 ex\hbox{$ < $}\kern -0.8 em
 \lower 0.7 ex\hbox{$\sim$}\,}
\def\gtsimeq{\,\raise 0.3 ex\hbox{$ > $}\kern -0.8 em
 \lower 0.7 ex\hbox{$\sim$}\,}
%
%
\def\rfnce{\par\noindent\hangindent 20pt {}}
\vbox to 0.1 in {}
\noindent
{\bf Protostellar fragmentation in a power-law density distribution$^*$}
\vskip 0.36 in\noindent
Andreas Burkert$^1$, Matthew R. Bate$^1$, and   Peter Bodenheimer$^2$

\noindent
$^1${\it Max-Planck-Institut f\"ur 
 Astronomie, K\"onigstuhl 17, D-69117 Heidelberg, Germany}

\noindent    
$^2${\it University of California Observatories/Lick Observatory,
Board of Studies in Astronomy and Astrophysics,
University of California, Santa Cruz, CA 95064, USA}

\vskip 0.2 in\noindent
Received 1996..............
\vskip 0.50 in\noindent
$^*$UCO/Lick Observatory Bulletin, No........
\np
\vglue 0.1 in
\noindent
{\bf ABSTRACT}

\noindent
Hydrodynamical calculations in three space dimensions 
of the collapse of an isothermal,
rotating 1 M\sol protostellar cloud 
are presented. The initial density stratification is a
power law with density $\rho \propto r^{-p}$,
with $p=1$. The case of the singular
isothermal sphere ($p=2$) is not considered; however
$p=1$ has been shown observationally to be a good representation
of the density distribution in
molecular cloud cores just before the beginning of collapse. 
The collapse is studied with  two independent numerical methods, 
an SPH code with 200,000 particles, and a finite-difference code with
nested grids which give high spatial resolution in the inner regions. 
Although previous numerical studies have indicated that such
a power-law distribution would not result in fragmentation into
a binary system, 
both codes show, in contrast, that    
multiple fragmentation does occur in the central regions of
the protostar. 
Thus the process of binary formation by fragmentation is
shown to be consistent with the fact that a large fraction of
young stars are observed to be in binary or multiple systems. 
\np\noindent
{\bf 1  INTRODUCTION }
\vskip 0.12 in\noindent
This paper considers a long-standing problem regarding binary
formation by fragmentation of a collapsing, rotating cloud, namely, that
observed 
molecular cloud cores in star forming regions are sufficiently
centrally condensed so that they would not fragment according to 
numerical calculations, yet these regions have a high proportion of
binary systems. 
Fuller \& Myers (1992), in a survey of cloud cores, some of 
which are in Taurus,  found a density distribution $\rho \propto r^{-p}$, where
$p \approx 1.6$, independent of whether the cores
had IR sources in them or not.  The survey by Ward-Thompson et al (1994), in the
dust continuum,  of cores 
without infrared sources showed $p \approx 1.25$ in the inner regions
and $p \approx 2.0$ in the outer regions, where the `inner' region 
refers to $r < 0.05$ pc.  One of these cores, L1689B, has been 
reobserved at 1.3 mm at higher spatial and angular resolution 
(Andr\'e, Ward-Thompson, \& Motte 1996) and the result has been confirmed: 
the central regions have $p \approx 1$. 
In addition, theoretical studies 
(Lizano \& Shu 1989; Tomisaka, Ikeuchi, \& Nakamura 1990; 
Ciolek \& Mouschovias 1994; Basu \& Mouschovias 1994) of the        
quasistatic evolution of magnetically supported molecular clouds end up
with cores having 
a centrally condensed distribution at the onset of collapse.
 
However, according to previous three-dimensional hydrodynamical 
calculations of the collapse of such cores (Boss 1987; Myhill \& Kaula 1992)
they do not fragment if they are initially  uniformly rotating
and have $p = 1$.  The same conclusion is reached for clouds with $p = 2$, 
an often-used initial condition for numerical calculations of star 
formation, and in this case the conclusion is bolstered by a linear
stability analysis (Tsai \& Bertschinger 1989). 
Although Myhill \& Kaula (1992) do find fragmentation with an initially 
differentially rotating cloud ($ p = 1 $ or 2), there is no observational evidence
that clouds at the onset of collapse are differentially rotating, and 
furthermore, it is expected from theory (Basu \& Mouschovias 1994) that
the initial clouds will be in near-uniform rotation because of the
effects of magnetic braking. 
  
From what has just been said one would
deduce, for example for the Taurus-Aurigae star-forming region, that very
few of the formed stars would be in binary systems. However in fact
almost all of the stars in that region are multiple, as deduced from the
observations of several groups (Ghez, Neugebauer, \& Matthews 1993; 
Leinert et al 1993; Simon et al 1995; Brandner et al 1996; Mathieu 1994). 
In other star-forming regions (Brandner et al 1996) the binary fraction 
is not quite so high as in Taurus, but nevertheless it is comparable
to that among nearby solar-type main-sequence stars (Duquennoy \& 
Mayor 1991). It is a generally accepted conclusion that most of
these binaries must have formed during the protostellar collapse, 
perhaps by fragmentation; Fuller, Ladd, \& Hodapp (1996) have recently
discovered a binary protostar in the earliest stages of formation.  

In the present paper we consider the question of whether 
fragmentation during protostellar collapse is, in fact, a viable
mechanism for binary formation.  Although many previous studies of
three-dimensional protostar collapse have resulted in fragmentation 
(Boss 1993, 1996;  Burkert \& Bodenheimer 1993,1996, Bonnell \& Bastien 1993, 
Bonnell \& Bate 1994, Monaghan 1994; Miyama 1992; 
Turner et al 1995; Nelson \& Papaloizou 1993; Sigalotti \& Klapp 1994), 
the initial assumed density variation with distance
from the centre  was generally either uniform
or a mild exponential (Boss 1991, 1993),that is, less 
centrally condensed than the distributions suggested by observations. 
We calculate the collapse of isothermal, rotating clouds, without
magnetic fields, with two different three-dimensional
hydrodynamic codes, starting with the density 
distribution $p = 1$
and with uniform rotation. The initial density distribution is 
assumed to be 
spherically symmetric with a small nonaxisymmetric perturbation. 
The purpose of our calculation is to calculate the collapse with  
much higher spatial resolution and to carry it over a longer time
period than earlier numerical calculations to see if fragmentation
does in fact occur.  We also test to see if fragmentation properties
depend on the assumed angular momentum of the cloud. 

\vskip 0.35 in\noindent
{\bf 2  THE COMPUTATIONAL METHOD AND INITIAL CONDITIONS}
\vskip 0.18 in\noindent
{\bf 2.1 Grid code}
\vskip 0.1 in\noindent 
The finite difference method is essentially the same as that described 
by Burkert \& Bodenheimer (1993, 1996). 
The calculations are performed on a 3-dimensional Eulerian, Cartesian
grid; the advection scheme is based on second-order monotonic
transport (van Leer 1977).
The full computational region is represented
by a standard grid, composed of 64$^3$ grid cells equally spaced in all
directions. For improved resolution of the 
inner regions,  four Cartesian    nested concentric 
subgrids were  
superimposed on the standard grid, giving a ratio of total cloud size 
to the size of the smallest zone of 1000 or 2000. 
The linear scale on a given subgrid is
reduced by a factor of 2 or 4 with respect to the next larger grid. 
The grid structure is set up at the beginning of the calculation and left
fixed during the entire run. 
An  artificial viscosity of the type described by 
Colella \& Woodward (1984) is added; its main effect is to suppress
a rapid increase in the density at centres of fragments.
Unlike the calculations of Burkert \& Bodenheimer (1996) there is no
symmetry assumed with respect to the $z$-axis, although the 
configuration is assumed to be symmetric with respect to the equatorial 
plane. 

An important difference between the calculations reported in this paper
and most previous fragmentation calculations is that an additional
numerical stability criterion is incorporated. Klein, Truelove,  \& 
McKee (1996) have pointed out that numerical fragmentation could
occur in a calculation if the local zone size is larger than the 
local Jeans length. In their adaptive mesh calculation they simply 
dynamically rezone to a sufficiently high level of accuracy so
that the criterion is always satisfied. In our calculations, to 
avoid using an excessively large number of zones, we introduce an 
artificial heating at high densities, so that if the mass of a 
zone exceeds the local Jeans mass, the zone is heated until its mass
is less than the local Jeans mass. The heating procedure affects 
only the regions of highest density, which in general are already
part of fragments, and tests of calculations with and without heating
show very little difference in the results. 

The Jeans mass is given by (Spitzer 1978) as $\rho l_J^3$ where
$l_J$ is the local Jeans length; thus 
$$ M_J = {2.44 \times 10^{23}} {({{T}\over 
{\mu}})^{3/2}}{\rho^{-1/2}}~~~{\rm g} \eqno(1)$$
If the zone size is $\Delta$, then the requirement that the zone mass
$\rho \Delta^3 < M_J$ defines a critical density $\rho_c$:
$$ \rho < \rho_c  = { {3.905 \times 10^{15} } \over {\Delta^2}}
{({{T}\over {\mu}})}~~~{\rm g~cm^{-3} } \eqno(2)$$
If $\rho > \rho_c$, then the zone is heated so that its new sound speed $c_s$ is
$$ c_s^2 = { {2.561 \times 10^{-16} \rho \Delta^2 R_g} \over 
{\eta_\rho}}~~~{\rm cm^{2}~s^{-2} } \eqno(3)$$
Here $R_g$ is the gas constant and $\eta_\rho$ is a safety factor, taken
to be 0.5, so that the heating starts at somewhat lower densities than 
$\rho_c$. 

\vskip 0.1in\noindent
{\bf 2.2 SPH code}




\vskip 0.1 in\noindent
The SPH calculations were performed using a 
three-dimensional code based on a version originally
developed by Benz (1990; Benz et al. 1990).
The standard form of artificial viscosity is used
(Monaghan \& Gingold 1983; Benz 1990; Monaghan 1992), 
with the parameters $\alpha_{\rm v}=1$ and $\beta_{\rm v}=2$.
The smoothing lengths of particles are variable in time and space,
subject to the constraint that the number of neighbours
for each particle
must remain approximately constant at $N_{\rm neigh}=50$.  
The SPH equations are integrated using a second-order Runge-Kutta-Fehlberg 
integrator.  

Two major modifications have been made to the original code.  First,
individual time steps are used for each particle (Bate, Bonnell \& Price 1995)
rather than a single time step for all particles.
The result is a great saving in computational time for simulations 
where there is a large density contrast (e.g. fragmentation calculations).  
Second, gravitational forces 
and a particle's nearest neighbours are found by using either 
a tree, as in the original code, or the special-purpose GRAvity-piPE
(GRAPE) hardware.  The implementation of SPH using the GRAPE
closely follows that described by Steinmetz (1996).
Using the GRAPE attached to a Sun SPARCstation 20 workstation typically 
results in a factor of 5 improvement in speed over the workstation alone.
The calculation presented here was performed using the GRAPE.

We use the polytropic equation of state 
$P = K \rho^{\gamma}$,
where $K$ is a constant that depends on the entropy of the gas.
The polytropic constant $\gamma$ varies with density as
$$
\hskip -0.15truecm \gamma = 1,~~~~~~~~\rho \leq \rho_{\rm c2} = 10^{-12} {\rm \ g\ cm}^{-3}, 
$$
$$
\hskip  -0.15truecm \gamma = {5 \over 3},~~~~~~~~\rho > \rho_{\rm c2} = 10^{-12} {\rm \ g\ cm}^{-3},
\eqno(4)$$
and $K$ is defined such that, when the gas is isothermal ($\rho \leq \rho_{\rm c2}$), $K=c_{\rm s}^2$ with the sound speed
$c_s = 2.037 \times 10^4 {\rm ~cm~s}^{-1}$.  Heating the gas 
when $\rho > \rho_{\rm c2}$ serves two purposes.
First, with a purely isothermal equation of state, a non-rotating 
fragment collapses to infinite density.  Thus, a fragmentation calculation 
must be stopped when the first fragment is formed.  
If the gas is heated, however, the collapse of a
fragment is halted and the calculation can be followed further.
Second, heating of the gas is required so that the minimum mass that can
be resolved ($\approx 2 N_{\rm neigh}$ times the particle mass) is always
less than a Jeans mass.  If this criterion is not met
fragmentation may be artificially induced (Bate \& Burkert, in preparation).
This criterion is similar  to that used for the grid code.  
For the calculation presented in this paper
200,000 equal-mass particles are used.  Thus, from equation 1, a Jeans mass is less
than the mass of $2 N_{\rm neigh}$ particles when 
$\rho \gtsimeq  7 \times 10^{-12} {\rm \ g\ cm}^{-3}$.  Heating the gas
for $\rho > \rho_{\rm c2}$ ensures a Jeans mass is always resolved.


 
 
 
 
 

 
\vskip 0.1in\noindent 
\vskip 0.1in\noindent 
{\bf 2.3 Initial conditions} 
\vskip 0.1in\noindent 
The protostar 
in the present calculation is assumed to be isothermal in space and time.
The initial conditions are specified by the ratios $\alpha$ and $\beta$, 
which are, respectively, the thermal and rotational energies divided
by the absolute value of the gravitational energy, and by the angular 
momentum distribution, the size and
form of the initial perturbation, and the form of the density
distribution. The latter is taken to be
$$ {\rho (r)} = {\rho_0}~{\left( {{r_0} \over {r}  }\right)^p}   \eqno(5)$$
with $p=1$. 
Here $r$ is the distance to the origin
and $\rho_0$ and $r_0$ are constants. The form of the perturbed profile is 
$$    {\rho_p (r)} = {\rho (r)}{ [1 + a_1 \cos ( 2 \phi)]}     \eqno(6)$$
where $a_1$ is the amplitude of the perturbation and $\phi$ is the
azimuthal angle
about the rotation ($z$) axis. In all  cases presented here the
$a_1$ is set to 0.1, the cloud mass $M_{tot}$ is set to 
1.0 M$_\odot$,  $r_0 \rho_0 = 0.1273$ g cm$^{-2}$,
the angular velocity $\Omega$ is uniform, and the 
radius $R$ of the original sphere is 5 $\times 10^{16}$ cm. 
The sound speed $c_s = 2.037 \times 10^4$ cm s$^{-1}$; 
thus the value of $\alpha$ for all calculations was 0.35.

\vskip 0.65 in\noindent
{\bf 3  RESULTS}
\vskip 0.18 in\noindent
{\bf 3.1 Analytic discussion}
\vskip 0.12 in\noindent 
A spherically symmetric, isothermal gas cloud with a power-law
density distribution as given by equation 5 is a somewhat unphysical
initial condition for $0 < p < 2$ because there exists a critical
radius $r_{crit} > 0$ inside which the gas will initially expand
even if centrifugal forces are neglected. Within $r_{crit}$ the
pressure gradient initially exceeds the gravitational force:

$$
- {   {c^2}\over {\rho }}{{ \partial \rho}\over { \partial r}} \geq 
{{G M(r)}\over {r^2}}
\eqno(7) $$
where $M(r)$ is the total mass inside $r$ and equality holds for
$r = r_{crit}$. Given the density distribution of equation 5 and
using equation 7, we find 

$$               
r_{crit} = R \left( {{2p(3-p)\alpha}\over {3(5-2p)}} \right)^
           {{1}\over {2-p}}
\eqno(8)       $$ 
which is valid for $0 \leq p < 2$.

Figure 1 shows $r_{crit}/R$ and the fractional mass of the expanding
region $M(r_{crit})/M_{tot}$ as a function of $p$ for different values of 
$\alpha $. Note that all power-law profiles with $0 < p <2$, even with
small values of $\alpha$, have an initially expanding core. 
For the present case, where $\alpha = 0.35$ and $p = 1$, 
$r_{crit}/R = 0.156$ and $M(r_{crit})/M_{tot} = 0.024$.   
\vskip 0.1 in\noindent
\vskip 0.18 in\noindent
{\bf 3.2 Fragmentation, grid code, high beta }          
\vskip 0.1 in\noindent
The first run was calculated with $\beta = 0.23$ 
on four  subgrids, each 128 $\times 128 \times 64$ zones in the 
($x, y,  z$) directions, respectively, which
had radii of 2.50 $\times 10^{16}$ cm, 1.25 $\times 10^{16}$ cm,
6.25  $\times 10^{15}$ cm, and 3.125 $\times 10^{15}$ cm. The resolution on the finest grid is 
4.88 $\times 10^{13}$ cm. 
Although the initial
central density was 6 $\times 10^{-15}$ g cm$^{-3}$  the initial 
expansion phase, which affects the inner 25\% of the radius, 
results in a reduction of the central density to about 2 $\times 10^{-17}$
 g cm$^{-3}$. The density and velocity distributions on the first 
subgrid at a time 4.3 $\times 10^{11}$ s are  shown in Figure 2. At 
this time the expanding central region has nearly reached its 
maximum extent, and is about to resume collapse. The density profile is
markedly less steep in this region than in the original cloud. The 
free-fall time in the central regions is now 4.4 $\times 10^{11}$ s. 

The collapse then proceeds through more than six orders of magnitude
increase in the central density before any signs of fragmentation
begin to appear. Figure 3 shows the development of the central region
(the innermost subgrid) starting at 1.176 $\times 10^{12}$ s from the
beginning of the calculation. In Figure 3a a definite central
density maximum is present. A small region around it has formed
a flattened disk structure and a spiral-arm 
pattern.  At the end points of the spiral pattern one can already
see the development of two condensations which accrete material
until they become
self-gravitating. Figure 3b shows the original central density
maximum connected by inner spiral arms  to the newly-formed
binary.  Outside this region an outer spiral pattern has developed
and a circum-triple disk is beginning to form. This high-density
disk has become well-developed a short time later (Fig. 3c). At that time
the symmetry of the inner system is broken as a
result of small numerical perturbations and the triple system rearranges itself
into a more stable hierarchical system, with a close binary and a third
object further out. At the outer edge of 
the disk two new density maxima are visible, triggered by the interaction
of the outer spiral arms. One of these maxima finally develops into an
outer fragment, the other one is tidally disrupted and subsequently accreted by
the outwards moving fragment of the previous inner triple system (Fig. 3d).
At $t = 1.217 \times 10^{12}$ s the masses of the inner binary components 
are 0.055 and 0.025 $M_{\odot}$. The mass of the third component of
the initial triple system is 0.017 $M_{\odot}$. The fourth fragment
has a mass of 0.011 $M_{\odot}$.

At this time we stop the calculation; on the inner grid (scale 
3 $\times 10^{15}$ cm) 90\% of the total mass is included in
the four fragments. However only a small fraction (12.4\%) of $M_{tot}$
is included in this region. We expect further interactions of
these fragments to take place and possibly further fragmentation
to occur on larger scales. But at this point it is clearly
established that fragmentation has occurred, that some of the
fragments will survive, and that the end result will be a 
multiple system.

\vskip 0.18 in\noindent
{\bf 3.3 Fragmentation, SPH code, high beta}        
\vskip 0.1 in\noindent

To test whether or not the fragmentation is dependent on the numerical 
method, we perform the same calculation as in Section 3.2, but use
Smoothed Particle Hydrodynamics (SPH).  We use 200,000 equal-mass particles
in order to resolve a Jeans mass at densities up to $10^{-12}$ g cm$^{-3}$
while maintaining an isothermal equation of state.  Beyond 
$10^{-12}$ g cm$^{-3}$ the gas is heated so that a Jeans mass is always 
resolved.

The early evolution is identical to that calculated by the grid code, with
an expansion in the centre of the cloud.
The initial resolution in the centre is not 
as good as with the grid code; the maximum
density is only $5 \times 10^{-16}$ g cm$^{-3}$.  This results from the fact
that, with SPH, the resolution depends on the density.
The smoothing length $h$, which gives the resolution, is defined so that
$2h$ contains $N_{\rm neigh}$ particles.  In the centre of the cloud, although
the density formally goes to infinity as the radius goes to zero, the mass 
fraction is very small, and hence $h$ must be large in order to 
contain $N_{\rm neigh}$ particles.  
Other than the difference in initial central density, however, the 
expansion is identical to that given by the grid code, with the same
density minimum of $2 \times 10^{-17}$ g cm$^{-3}$ being reached before 
the expansion stops and the collapse begins.

During the collapse, the central density increases by more than
six orders of magnitude, in agreement with the grid code.  A central
density maximum forms, and a disk of gas begins to form around it.
The evolution of the central density maximum and its disk is shown in
Figure 4.  As soon as the disk appears around the central density
maximum it is threaded by spiral arms (Fig. 4a).  The arms continually
wrap-up, interact and reform because of their differential rotation.
Ultimately, density maxima form at the ends of the two arms leading to
a binary around the central object (Fig. 4b).  As higher
angular momentum material falls in, spiral arms extend beyond the binary
into a circum-triple disk.  The symmetry of the triple system rapidly
breaks, causing the outer spiral arms to become asymmetric.  
Again, density maxima
form at the ends of these arms, however, due to the asymmetry, 
only one of these forms a fragment immediately; the other is temporarily
disrupted (Fig. 4c).  Eventually, this disrupted fifth density maxima does
gather enough material to form a fifth fragment (Fig. 4d). Note that
a fifth fragment appears also temporarily in the grid code. There, however,
the somewhat different orbital evolution of the inner triple system leads
to a merger (Fig. 3d).

The collapse and fragmentation follows the same pattern as with the grid code:
a central object is formed, surrounded by a disk with spiral arms;
fragments form at the end of the arms; the initial symmetry of the triple
system breaks with one fragment moving outwards and the other two components
forming a close binary system;  a circum-triple disc forms,
again threaded by spiral arms which fragment, with the details 
depending on symmetry breaking of the triple.  
The fact that this pattern of fragmentation
has been produced with two completely different hydrodynamic codes 
demonstrates that the fragmentation is physical and not a numerical effect.
Some details of the fragmentation differ between the two codes.  
The formation of the triple system occurs slightly earlier with the SPH code 
($t=1.155 \times 10^{12}$ s) than the grid code ($t=1.176 \times 10^{12}$ s), 
before the disc has reached as large an extent.  
This earlier fragmentation is probably due to the 
different resolutions of the two codes, and the 
higher numerical noise that is present in the SPH code.
The resolution of the two codes are almost identical at densities of
$1.0 \times 10^{-12}$ g cm$^{-3}$ ($h=4.0 \times 10^{13}$ cm, 
while the smallest grid cell has $\Delta=4.9 \times 10^{13}$ cm).  
However, because the 
resolution of SPH depends on the density, for $\rho > 10^{-12}$ g cm$^{-3}$ 
(i.e. in the spiral arms) the resolution is actually higher with SPH.  
The higher numerical noise of the SPH code results from
the use of the GRAPE hardware which has less than single precision, 
whereas the grid code calculation (using a CRAY) performs 
gravitational calculations with double precision.
This difference in numerical noise also leads to the symmetry of the triple
system being broken quicker with the SPH code than the grid code.

We stop the calculation at the time $1.190 \times 10^{12}$ s (Fig. 4d).  
At this point the disk and fragments contain 12.4\% of $M_{tot}$.
The fragment masses are 0.050 and 0.020 $M_{\odot}$ for the close binary and
0.018 $M_{\odot}$ for the third component of the initial triple system.
These values are in excellent agreement with the results of the grid code. 
The two fragments which formed later on the far-left and far-right of Figure 4d
have masses of 0.017 $M_{\odot}$ and 0.003 $M_{\odot}$, respectively.
As with the grid code, 
mergers between fragments and further fragmentation are probable
as the remaining material falls in, but the survival of a multiple system
is likely.

\vskip 0.18 in\noindent
{\bf 3.4 Fragmentation, grid code, low beta}
\vskip 0.1in\noindent 
The runs discussed in the previous sections were  calculated 
with fairly large rotation ($\beta = 0.23$).   In order to check
whether such a high value of $\beta$ is crucial  for fragmentation
we have recalculated the same case with $\beta$ reduced by a factor of 2. 
As the gas is now less rotationally supported we expect
the system to become more centrally condensed. We therefore 
increased the resolution on the innermost grid by a factor of
two, compared with the grid setup used in section 3.2.
The four  subgrids, each 128 $\times 128 \times 64$ zones in the 
($x, y,  z$) directions, respectively, now have
radii of 2.50 $\times 10^{16}$ cm, 6.25 $\times 10^{15}$ cm,
3.125 $\times 10^{15}$ cm, and 1.563 $\times 10^{15}$ cm. 
The resolution on the finest grid is 2.44 $\times 10^{13}$ cm. 

Figure 5 shows snapshots of the evolution, which is
qualitatively the same as for the high-$\beta $ case, except that
the scale of the region of initial fragmentation is reduced by a factor of 2.
The initial development of spiral arms with density maxima at the
ends is shown in Figure 5a. Figure 5b shows an inner
triple system which formed in an analogous manner to the systems shown
in Figures 3b and 4b.
The symmetry then breaks as the original central density maximum 
approaches one of the new fragments, and a fourth fragment
appears near the outer edge of the disk (Fig. 5c). A short time later 
five fragments are present (Fig. 5d) and the surrounding circum-fragment
disk has expanded considerably.  The calculation is stopped before 
the disk has grown to the size and mass where induced fragmentation
on a larger scale (such as that shown in Fig. 3d) would be expected. 
However we obtain again multiple fragmentation;
five distinct fragments, including one of very low mass,
have clearly been established (Fig. 5d.)
The typical separations are now of order $10^{15}$ cm as compared with 
$4 \times 10^{15}$ cm at the end of the 
previous runs. The masses of the fragments
in Figure 5d are (from left to right) .016, .017, .03, .002, and 
.016 $M_{\odot}$, slightly lower in total than in the previous case 
but including practically the entire mass on the innermost subgrid 
(scale 1.56 $\times 10^{15}$ cm).
Note also that in this case the timescale of fragmentation is
slightly shorter. This case was not run further because at $t = 1.007 
\times 10^{12}$ s 
it is clear that fragmentation does occur even with the lower value of $\beta $.

The fragmentation is very likely not to be of numerical origin, since it
all takes place on the innermost grid and since heating is included to 
keep zone masses below the Jeans mass.  In fact, heating plays only a
very minor role in the sequence of events. From equation (2), the critical
density above which heating takes place is 
$\rho_c = 3.27 \times 10^{-11}$ g cm$^{-3}$. The densities at the centres of
the fragments in Figure 5d are typically 1.6 $\times 10^{-10}$ g cm$^{-3}$.
and $\rho > \rho_c$ in only a few zones near the centres of the 
established fragments. 
The temperature increases to 98 K at the highest  densities, which of course
would tend to suppress fragmentation, but only in material 
which has already fragmented. The heating effect thus just slows the 
increase in the central densities. However the density of typical
material in the vicinity of the fragments, which is the same as that
from which the fragments originally condense, is about 10$^{-12}$ 
g cm$^{-3}$. There the heating has no effect.                      
\vskip 0.35 in\noindent
{\bf 4  CONCLUSIONS}
\vskip 0.18 in\noindent
The  main purpose of this paper has been to demonstrate that
multiple fragmentation occurs even from an initial power-law
density distribution which is similar to what is observed
in molecular cloud cores. Our results differ from those
of previous numerical simulations which, for initially
uniform rotation, did not show evidence for fragmentation.
However the initial expansion phase, which is an important
factor as it leads to central expansion and to a core with a flat
density distribution, had previously not been resolved.
Thus in fact the actual initial conditions for collapse are
unlikely to have $\rho \propto r^{-p}$ where $0 \le p \le 2$
but rather a distribution in which the central regions have a 
flatter profile. If such a configuration has solid-body rotation
its angular momentum distribution will be somewhat different
from that used here, but it is just as likely to fragment,
as has been demonstrated by the calculations, for example,
of Boss (1993).

In any case, high enough spatial resolution
is required to bring out the detailed structure of the central
regions. In addition, the calculations have to be carried out
for a sufficient time beyond that when a 
condensed central object has formed so that a sufficient amount of mass with
higher angular momentum has accumulated in the central disk region.
Our results show that at least three to four fragments with unequal
masses form, with the details depending on the value of $\beta$ and the 
exact time of symmetry breaking which depends on the numerical noise.

It is promising that two entirely independent numerical techniques lead
basically to the same result: the initial formation of a central triple
system, subsequent symmetry breaking,  and the formation of an outer disk
with additional fragmentation on a 
larger scale. As the further evolution of the system
is expected to be highly chaotic it will depend on the numerical technique and
the local resolution which are not exactly identical in the two codes.
It still has to be investigated by longer-term calculations to what extent
the final stage will depend on all these details. Another important question
is whether multiple fragmentation will occur in the extreme case of an
$p=2$ power law which does not lead to an initially expanding core.

\medskip 
\noindent
{\bf ACKNOWLEDGMENTS}
\vskip 0.18 in\noindent
This work was supported in part through National Science Foundation
grant AST-9315578 and in part through a special NASA astrophysics theory
program which supports a joint Center for Star Formation Studies at
NASA/Ames Research Center, UC Berkeley, and UC Santa Cruz.
AB thanks the staff of UCO/Lick Observatory for the hospitality 
during his visit, and PB thanks the staff of the Max Planck Institute for     
Astronomy in Heidelberg  for the hospitality during his visit. 

\medskip 
\noindent
{\bf REFERENCES}
\medskip
\rfnce{
 Andr\'e, P., Ward-Thompson, D., \& Motte, F. 1996, preprint 
}\rfnce{
 Basu, S., Mouschovias, T. Ch., 1994, ApJ, 432, 720            
}\rfnce{
 Bate M. R., Bonnell I. A., Price N. M., 1995, MNRAS, 277, 362
}\rfnce{
 Benz W., 1990, in Buchler J. R., ed., The Numerical Modeling of Nonlinear Stellar Pulsations: Problems and Prospects. Kluwer, Dordrecht, p. 269
}\rfnce{
 Benz W., Bowers R. L., Cameron A. G. W., Press W., 1990, ApJ, 348, 647
}\rfnce{
 Bonnell, I., Bastien, P., 1993, ApJ, 406, 614                   
}\rfnce{
 Bonnell, I., Bate, M. R., 1994, MNRAS, 269, L45                 
}\rfnce{
 Boss, A. P., 1987, ApJ, 319, 149 
}\rfnce{
 Boss, A. P., 1991, Nat, 351, 298 
}\rfnce{
 Boss, A. P., 1993, ApJ, 410, 157 
}\rfnce{
 Boss, A. P., 1996, ApJ, 468, 231 
}\rfnce{
 Brandner, W., Alcala, J. M., Kunkel, M., Moneti, A., Zinnecker, H., 1996, 
A\&A, 307, 121 
}\rfnce{
 Burkert, A., Bodenheimer, P.,  1993, MNRAS, 264, 798 
}\rfnce{
 Burkert, A., Bodenheimer, P.,  1996, MNRAS, 280, 1190
}\rfnce{
 Ciolek, G. E.,  Mouschovias, T. Ch., 1994, ApJ, 425, 142 
}\rfnce{
 Colella, P., Woodward, P., 1984, J. Comput. Phys. 54, 174 
}\rfnce{
 Duquennoy, A., Mayor, M., 1991, A\&A, 248, 485
}\rfnce{
 Fuller, G. A., Ladd, E. F., \& Hodapp, K.-W., 1996, preprint
}\rfnce{
 Fuller, G. A., Myers, P. C., 1992, ApJ, 384, 523            
}\rfnce{
 Ghez, A. M., Neugebauer, G.,  Matthews, K., 1993, AJ, 106, 2005 
}\rfnce{
 Klein, R., Truelove, K., \& McKee, C., 1996, personal communication
}\rfnce{
 Leinert, C., Zinnecker, H., Weitzel, N., Christou, J., Ridgway, S. T., 
Jameson, R., Haas, M., Lenzen, R., 1993, A\&A, 278, 129
}\rfnce{
 Lizano, S.,  Shu, F., 1989, ApJ, 342, 834
}\rfnce{
 Mathieu, R. D., 1994, Annu. Rev. Astron. Astrophys, 32, 465            
}\rfnce{
 Miyama, S. M., 1992, Publ. Astron. Soc. Japan, 44, 193            
}\rfnce{
 Monaghan J. J., 1992, ARA\&A, 30, 543
}\rfnce{
 Monaghan, J. J., 1994, ApJ, 420, 692 
}\rfnce{
 Monaghan J. J., Gingold R. A., 1983, J. Comput. Phys., 52, 374
}\rfnce{
 Myhill, E.,  Kaula, W. M., 1992, ApJ, 386, 578
}\rfnce{
 Nelson, R. P., Papaloizou, J. C. B., 1993, MNRAS, 265, 905 
}\rfnce{
 Sigalotti, L. DiG., Klapp, J., 1994, MNRAS, 268, 625 
}\rfnce{
 Simon, M., Ghez, A. M., Leinert, Ch., Cassar, L., Chen, W. P., Howell, R. R., 
 Jameson, R. F., Matthews, K., Neugebauer, G., Richichi, A., 1995, 
ApJ, 443, 625 
}\rfnce{
 Spitzer, L. Jr. 1978, Physical Processes in the Interstellar Medium, 
Wiley, New York 
}\rfnce{
 Steinmetz M., 1996, MNRAS, 278, 1005
}\rfnce{
 Tomisaka, K., Ikeuchi, S.,  Nakamura, T., 1990, ApJ, 362, 202 
}\rfnce{
 Tsai, J,, Bertschinger, E., 1989, Bull. Am. Astron. Soc., 21, 1089 
}\rfnce{
 Turner, J. A., Chapman, S. J., Bhattal, A. S., Disney, M. J., Pongracic, H., 
\& Whitworth, A. P. 1995, MNRAS, 277, 705
}\rfnce{
 van Leer, B., 1977, J. Comput. Phys., 23, 276
}\rfnce{
 Ward-Thompson, D., Scott, P. F., Hills, R. E.,  Andr\'e, P., 1994,
 MNRAS, 268, 276 
}

\vskip 0.2 in
\centerline{FIGURE CAPTIONS}

\vskip 0.1 in\noindent
{\bf Figure 1.} The radius r$_{crit}$, inside of which a spherical 
power-law cloud initially expands, is shown (left panel), in units of the
total cloud radius, as a function of the power-law index $p$, 
for various values of $\alpha$, the initial ratio of thermal 
energy to the absolute value of the gravitational energy. 
The right panel shows the mass fraction inside r$_{crit}$, as a function
of the same two parameters. 
From top to bottom, the curves correspond to $\alpha = 0.5, 0.4, 0.3, 
0.2, {\rm and}~ 0.1.$
\vskip 0.07 in\noindent
{\bf Figure 2.}  Contours of equal density in the equatorial ($x,y$)  plane
at $z = 0$
at a time of 4.3 $\times 10^{11}$ s in the high-beta grid-code
calculation. The first subgrid is shown. The maximum density is 
log $\rho_{max} = -16.67$ and the contour interval is $\Delta$ log 
$\rho = 0.03$. Velocity vectors are shown with length proportional 
to speed; the maximum velocity $V_{max}$ is 3.58 $\times 10^4$ cm s$^{-1}$.
The linear scale is given in cm. 

\vskip 0.07 in\noindent
{\bf Figure 3.} Evolution of the high-beta case with the grid code.
The innermost grid is shown. 
Symbols and curves have the same meaning as in Fig. 2. 
(a; upper left) t = 1.176 $\times 10^{12}$ s; log $\rho_{max} = 
-10.3$; $\Delta$ log  $\rho = 0.166$; $V_{max} = 1.95 \times 10^5$ cm s$^{-1}$.
(b; upper right) t = 1.190 $\times 10^{12}$ s; log $\rho_{max} = -10.2$; $\Delta$ log  $\rho = 0.17$; $V_{max} = 2.57 \times 10^5$ cm s$^{-1}$. 
(c; lower left) t = 1.209 $\times 10^{12}$ s; log $\rho_{max} = -10.1$; $\Delta$ log  $\rho = 0.173$; $V_{max} = 3.17 \times 10^5$ cm s$^{-1}$.
(d; lower right) t = 1.217 $\times 10^{12}$ s; log $\rho_{max} = -10.1$; $\Delta$ log  $\rho = 0.173$; $V_{max} = 3.67 \times 10^5$ cm s$^{-1}$.
\vskip 0.07in\noindent 
{\bf Figure 4.} Evolution of the high-beta case with the SPH code.  Symbols and
curves have the same meaning as in Fig. 2.  
(a; upper left) t=$1.135 \times 10^{12}$ s; log $\rho_{max} = -10.0$;
$\Delta$ log $\rho=0.25$; $V_{max}=2.44 \times 10^5$ cm s$^{-1}$.
(b; upper right) t=$1.160 \times 10^{12}$ s; log $\rho_{max} = -9.7$;
$\Delta$ log $\rho=0.25$; $V_{max}=2.88 \times 10^5$ cm s$^{-1}$.
(c; lower left) t=$1.175 \times 10^{12}$ s; log $\rho_{max} = -9.6$
$\Delta$ log $\rho=0.25$; $V_{max}=3.66 \times 10^5$ cm s$^{-1}$.
(d; lower right) t=$1.190 \times 10^{12}$ s; log $\rho_{max} = -9.5$;
$\Delta$ log $\rho=0.25$; $V_{max}=3.38 \times 10^5$ cm s$^{-1}$.
\vskip 0.07in\noindent
{\bf Figure 5.} Evolution of the low-beta case with the grid code.
The innermost grid is shown. 
Symbols and curves have the same meaning as in Fig. 2. 
(a; upper left) t = 9.637 $\times 10^{11}$ s; log $\rho_{max} = 
-9.65$; $\Delta$ log  $\rho = 0.219$; $V_{max} = 1.98 \times 10^5$ cm s$^{-1}$.
(b; upper right) t = 9.706 $\times 10^{11}$ s; log $\rho_{max} = 
-9.498$; $\Delta$ log  $\rho = 0.224$; $V_{max} = 1.88 \times 10^5$ cm s$^{-1}$.
(c; lower left) t = 9.840 $\times 10^{11}$ s; log $\rho_{max} = 
-9.46$; $\Delta$ log  $\rho = 0.225$; $V_{max} = 3.01 \times 10^5$ cm s$^{-1}$.
(d; lower right) t = 1.007 $\times 10^{12}$ s; log $\rho_{max} = -9.431$; 
$\Delta$ log  $\rho = 0.226$; $V_{max} = 3.41 \times 10^5$ cm s$^{-1}$. 
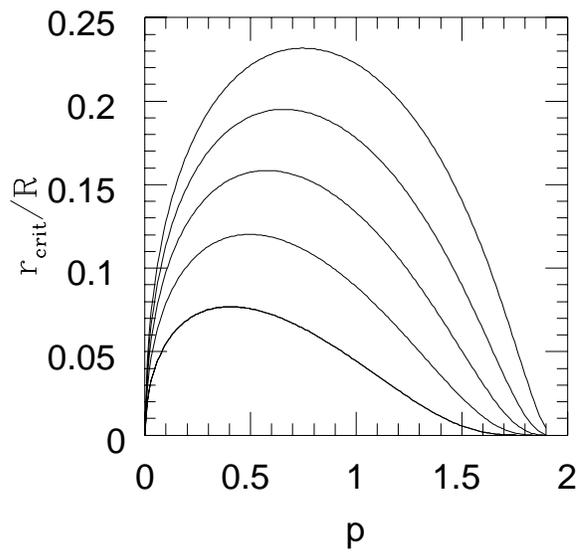
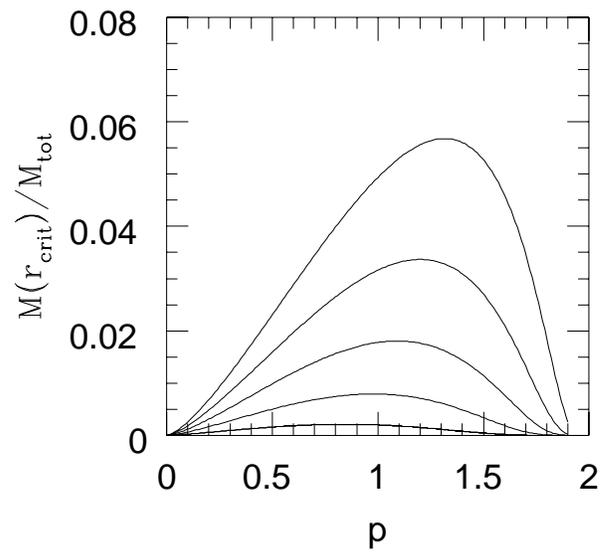
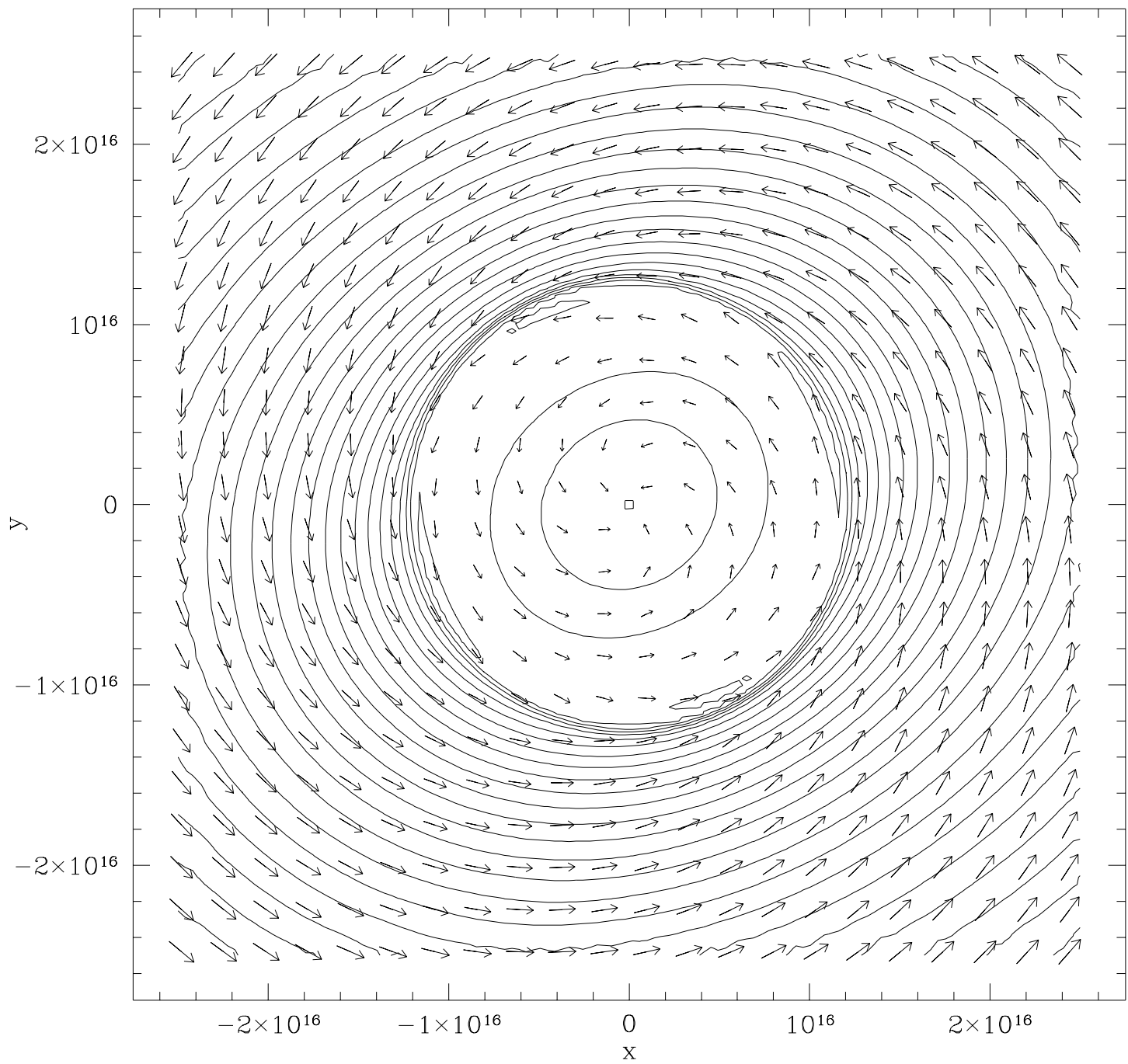
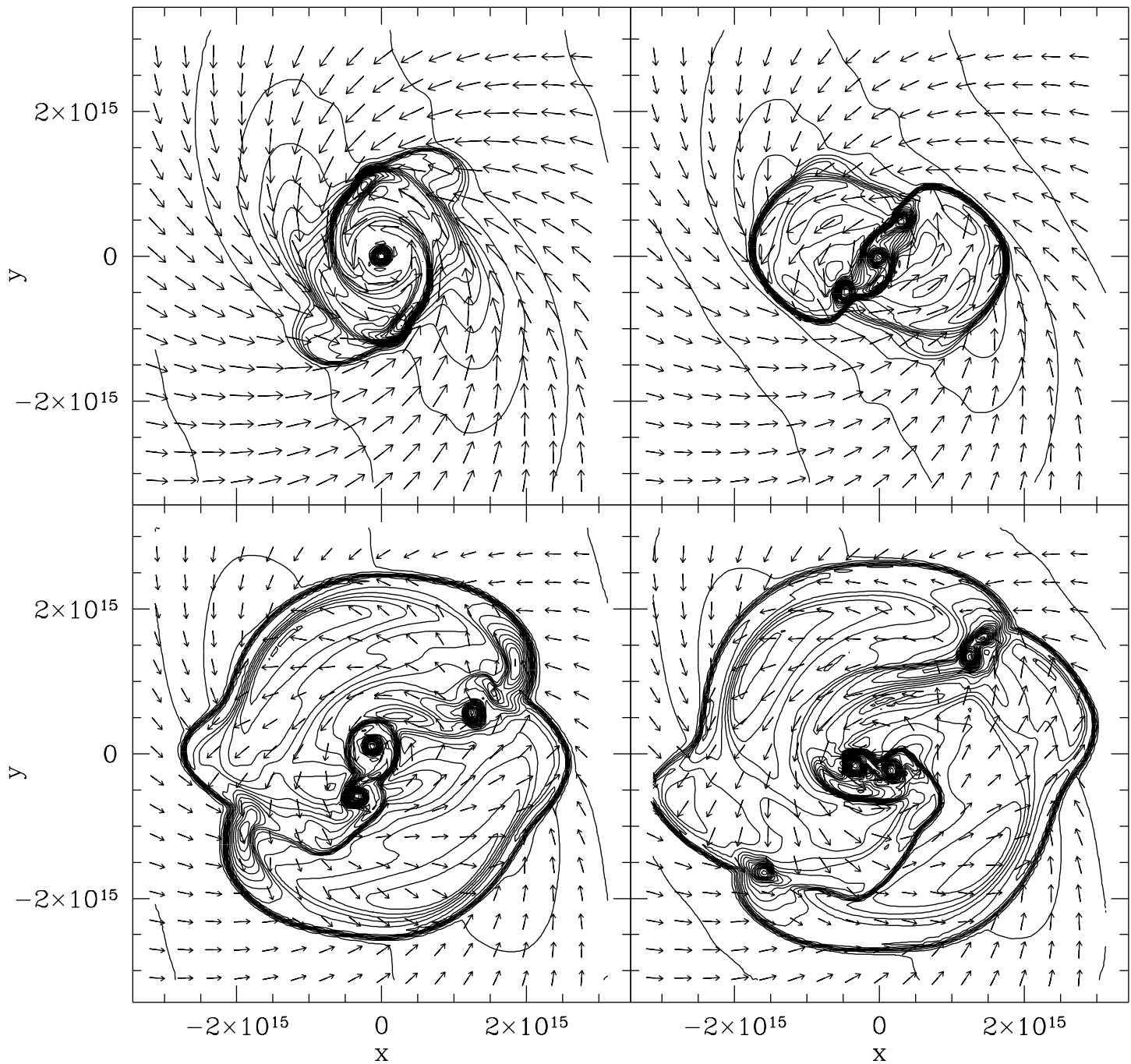
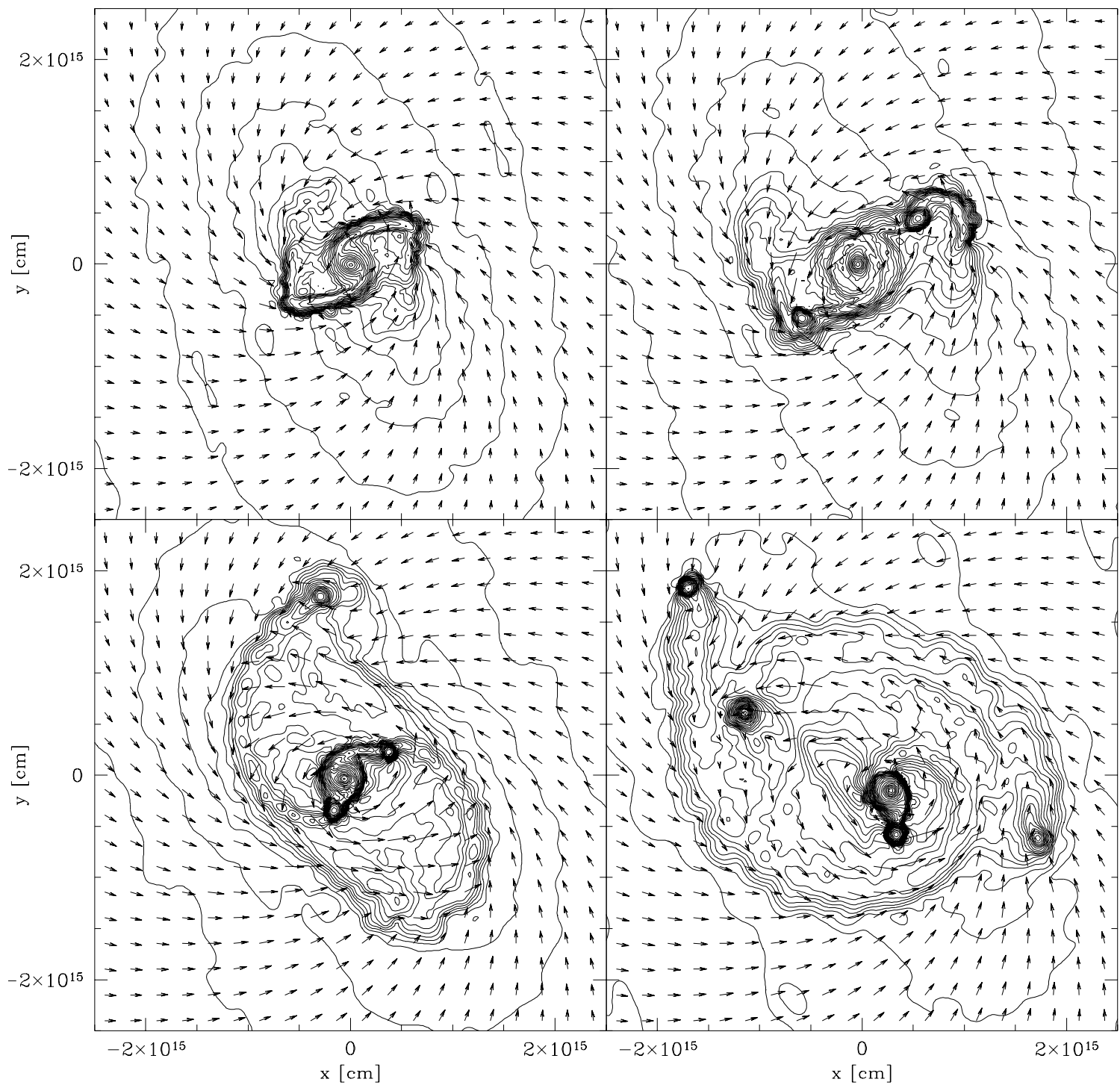
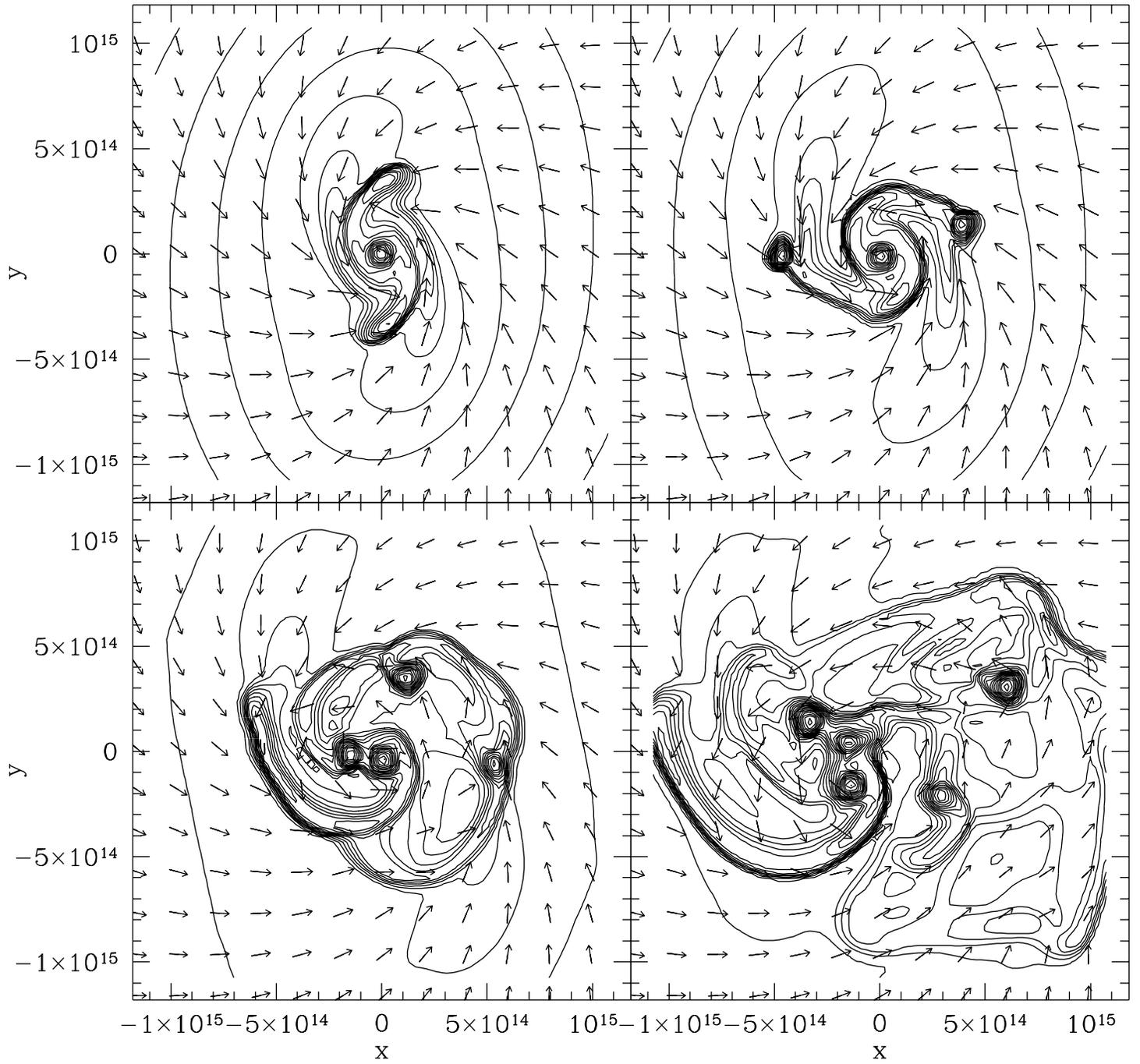
\bye